# DETECTING AN EAVESDROPPER IN QKD WITHOUT PUBLIC BIT COMPARISON

*S.M. Barnett*[1] *& S.J.D. Phoenix*[2]

[1]Dept of Physics, Strathclyde University, Glasgow, UK
[2]Khalifa University, Abu Dhabi Campus, PO Box 127788, Abu Dhabi, UAE

**Abstract**

*We present a method for determining the presence of an eavesdropper in QKD systems without using any public bit comparison. Alice and Bob use a duplex QKD channel and the bit transport technique for relays. The only information made public is the respective basis choices which must be revealed in standard QKD systems anyway. We find that every filtered bit can be used to determine the presence of errors without compromising the security. This is an improvement on using a random sample in the standard BB84 protocol.*

1. **Comparison of Bits in the BB84 Protocol**

Quantum Key Distribution (QKD) is a method for establishing a secret key between two parties, conventionally labelled as Alice and Bob, in which the laws of physics guarantee the security. The original technique, known as the BB84 protocol, works by exploiting quantum complementarity to encode information in the eigenstates of complementary operators. By randomly selecting which of these operators is used to encode the information an eavesdropper cannot retrieve that information without disturbing the integrity of some of the transmitted states. Other protocols have since been developed, but all rely on either complementarity or quantum correlation to ensure security. For an excellent review of the first decade or so of QKD see [1]. We focus here on the BB84 protocol, but with suitable adaptation, the technique can be extended to others.

Let $\hat{X}$ and $\hat{Y}$ be the operators representing complementary bases and consider a spin system consisting of just two orthonormal states. The eigenstates of the operators $\hat{X}$ and $\hat{Y}$ are $|\pm\rangle_X$ and $|\pm\rangle_Y$, respectively. The '+' states are given a bit value of 1 and the '−' states given a bit value of 0. Thus, there are two coding schemes which we can label with *X* and *Y*, respectively, in which each transmitted quantum state conveys a single bit. In the BB84 protocol the bit values associated with the quantum states are used to establish the key. The basic BB84 protocol can be outlined as follows

1. Alice chooses a coding basis at random
2. Alice chooses a state from that basis at random
3. For each well-defined timeslot Alice transmits her chosen quantum state to Bob
4. Alice records the triple (T, CB, BV) where T is the timeslot, CB is the coding basis and BV is the bit value.
5. For each well-defined timeslot Bob chooses, at random, a coding basis in which to read (i.e. measure) the transmitted quantum state
6. Bob records the triple (T, CB, BV)
7. Alice and Bob compare T and CB and discard (a) those timeslots in which Bob received no state and (b) those timeslots in which their choices of CB differ.

8. Alice and Bob now have a list in which T and CB are in agreement. A random sample of these results is chosen and the value BV compared. This gives an estimate of the error rate for the channel. These compared timeslots are discarded.
9. The remaining timeslots can be renumbered for convenience so that T = 1,2,3,4, . . .,.N where N is the total number of successful timeslots.

Step 6 in this protocol represents the end of the quantum transmission and measurement processes. The subsequent steps are to do with the processing of the results of the transmission and measurement.

In order to ensure that Alice and Bob can match those timeslots for which they should have the same BV they must compare their (T, CB) values. In the standard BB84 protocol either Alice or Bob (or both) simply publishes their list for the (T, CB) values they used. The other party can then compare the two lists and both parties discard any data for which the CB values disagree. For an ideal system in the absence of an eavesdropper this procedure ensures that Alice and Bob end up with an identical list (T, CB) which is, on average, about half the size of the original. In order to detect the presence of errors in this data Alice and Bob select a random sample (step 8 above) and compare the actual bit values BV.

## 2. Detecting the Presence of an Eavesdropper without Bit Comparison

In order to detect errors without any public bit comparison Alice and Bob must operate a duplex QKD channel. That is Alice transmits photons to Bob, according to the BB84 protocol, and Bob transmits photons to Alice according to the BB84 protocol. For convenience we shall imagine these to be interleaved so that for odd timeslots a photon is transmitted by Alice whereas Bob transmits in even timeslots. These can, in fact, be two entirely separate transmissions; all that is required is that we can uniquely correlate a particular transmission with a particular measurement. This uses the 'bit transport' technique developed in [2] where it was used to show how intercept/re-send relays can be used to extend the distance of QKD. The basic idea is that separate 'good' channels are correlated by linking different timeslots. The overall effective number of channels that can be used for key exchange is unaffected by the introduction of extra relays. We can view the current duplex channel as of the form *Alice – Relay – Bob* folded back on itself.

An example of such an interleaved transmission is shown in Table 1. We assume that each timeslot is occupied and that the photon reaches its destination. This is, of course, not true in practice, but it is easy to accommodate timeslots where nothing is transmitted or received.

| T | 1 | 2 | 3 | 4 | 5 | 6 | 7 | 8 | 9 | 10 | 11 | 12 | 13 | 14 | 15 | 16 | 17 | 18 | 19 | 20 |
|---|---|---|---|---|---|---|---|---|---|----|----|----|----|----|----|----|----|----|----|----|
| **ACB** | X |   | X |   | Y |   | X |   | Y |   | X |   | Y |   | Y |   | Y |   | X |   |
| **ABV** | 1 |   | 1 |   | 1 |   | 0 |   | 1 |   | 1 |   | 0 |   | 1 |   | 0 |   | 0 |   |
| **BCB** | Y |   | X |   | Y |   | Y |   | Y |   | X |   | X |   | Y |   | Y |   | Y |   |
| **BBV** | ◉ |   | 1 |   | 1 |   | ◉ |   | 1 |   | 1 |   | ◉ |   | 1 |   | 0 |   | ◉ |   |
| **BCB** |   | X |   | X |   | Y |   | X |   | X |   | Y |   | Y |   | Y |   | X |   | Y |
| **BBV** |   | 0 |   | 0 |   | 1 |   | 1 |   | 1 |   | 0 |   | 0 |   | 1 |   | 0 |   | 0 |
| **ACB** |   | X |   | Y |   | Y |   | X |   | Y |   | X |   | Y |   | Y |   | X |   | X |
| **ABV** |   | 0 |   | ◉ |   | 1 |   | 1 |   | 1 |   | ◉ |   | 0 |   | 1 |   | 0 |   | ◉ |

Alice informs Bob of her basis choices for both transmission and measurement. Bob filters this data into 3 sets. The first set is the data for which they expect no agreement because they have chosen different bases. In Table 1 this set consists of timeslots 1,4,7,12,13,19 and 20. Bob informs Alice of these timeslots and they both discard this data. The second set consists of those remaining timeslots in which Alice is the transmitter and in Table 1 consists of the timeslots 3,5,9,11,15 and 17. The third set is the remaining data and represents data in which Bob has initiated the transmission and Alice has measured in the correct basis. In Table 1 this set comprises the timeslots 2,6,8,10,14,16 and 18.

In a perfect world, and in the absence of an eavesdropper, Alice and Bob should agree on the data tagged by the timeslots. In order to check this Bob chooses a timeslot from set 2 and reads the value of BV he measured. He then looks for an element of set 3 with the same BV. He sends Alice the two timeslots. Alice compares the value of BV she has for these two timeslots. They should be equal. If there are errors on the channel, either caused by an eavesdropper or by practical imperfections, then there is a finite probability that Alice's comparison will fail. If Bob transmits a sufficient number of these timeslot pairs from set 2 and set 3 the probability of error remaining undetected can be made negligibly small.

If Bob transmits an extra bit along with each timeslot then the necessity of searching for an identical BV from set 3 to match the one from set 2 is unnecessary. The extra bit tells Alice whether to perform a bit flip on the bit from set 3 or not. Using this extra bit means that the elements of set 2 and set 3 can be used in sequence if so desired. Let's see how this works by using the example transmission of Table 1.

In Table 2 we have written BV for each timeslot and separated the data into the two sets, as discussed above.

| Set 2 (Alice's Transmission) | |
|---|---|
| Timeslot | BV |
| 3 | 1 |
| 5 | 1 |
| 9 | 1 |
| 11 | 1 |
| 15 | 1 |
| 17 | 0 |
| – | – |

| Set 3 (Bob's Transmission) | |
|---|---|
| Timeslot | BV |
| 2 | 0 |
| 6 | 1 |
| 8 | 1 |
| 10 | 1 |
| 14 | 0 |
| 16 | 1 |
| 18 | 0 |

After the communication Bob would send Alice a list of triples. The first value is from set 2, the second from set 3, and the extra bit tells Alice whether to flip the bit from set 3 with 1 taken to mean to perform the flip and 0 taken to mean leave alone. Although, for simplicity here we assume the rule that the flip is associated with set 3, all that is required is that one of the bits from set 2 or set 3 is flipped by Alice. An alternative way of viewing this is to think of the extra bit as a parity check. For the above example Bob would send Alice the following list of triples

(3,2,1)
(6,5,0)
(9,8,0)
(11,10,0)
(15,14,1)
(17,16,1)

The first triple, for example, tells Alice to compare her values of BV for timeslots 2 and 3 with the value of BV for timeslot 2 flipped. The comparison by Alice will only fail if there has been an error on the channel, either from Alice to Bob, or from Bob to Alice.

It is important to note that an eavesdropper does not gain any extra information from the extra bit that Bob transmits. It is not a parity bit in the usual sense of the meaning. The secret values of BV are never revealed. An eavesdropper will know which timeslots are considered legitimate candidates for forming a key from the basis information, but she knows this in the standard BB84 protocol too. She does, however, gain *some* extra information from knowing that the bit values in two timeslots are the same which she can infer from the timeslots in the list of triples transmitted by Bob. This gives her precisely one bit of extra information out of the two possible in these timeslots. She will know that timeslots 2 and 3 have bit values of 0,1 or 1,0, but not which of these is correct.

In order to eliminate this information gain by Eve, Alice and Bob could adopt the rule that the compared timeslots must be understood as follows. If the bit values are 0,1 or 0,0 then this is read as a 0. If the bit values are 1,0 or 1,1 then this is read as a 1. Alice and Bob reduce their potential key size by a factor of 2, but as it is a duplex channel this amounts to the same key potential key size as for the single channel BB84 protocol. So in the above example, the first triple sent by Bob is to be understood as having the final bit value 1.

The fundamental difference between this protocol and BB84 is that the duplex channel and bit transport mechanism allows Alice and Bob *to use their entire filtered transmission to check for errors* rather than just use a random sample which must then be discarded.